# Microfinance in Thailand: Navigating Challenges and Unlocking Opportunities

Worrawoot Jumlongnark[1*]

**Abstract**

This review article explores the challenges and opportunities faced by the Bank for Agriculture and Agricultural Cooperatives (BAAC) in Thailand from a microfinance perspective. It examines the role of BAAC as a specialized financial institution in assisting underprivileged households and small businesses in accessing financial services. The study emphasizes the challenges and opportunities faced by BAAC in promoting sustainable development. It also explores BAAC's role in advancing the BCG Model policy, which fosters sustainability in the agricultural sector through Bio Economy Credit, Circular Economy Credit, and Green Credit. These initiatives support investments in biotechnology, waste reduction (Zero Waste), organic farming, and safe food production, all aimed at enhancing farmers' quality of life, stimulating growth in agriculture, and preserving the environment. Moreover, BAAC remains committed to upholding transparency, fairness, and operational standards.

**Keywords:** Bank for Agriculture and Agricultural Cooperatives, microfinance institution, sustainable development.

**Introduction**

Specialized finance institutions play a crucial role in the global financial landscape, catering to specific sectors and addressing unique financing needs. These institutions vary across countries and regions, offering specialized services and products to support economic development, promote social welfare, and facilitate targeted financial inclusion (Bank of Thailand, 2023). They often operate alongside traditional commercial banks and provide financial solutions tailored to the requirements of specific industries, communities, or disadvantaged groups. Examples of specialized finance institutions include development banks, microfinance institutions, agricultural banks, housing finance institutions, export-import banks, and industrial development corporations. Development banks such as the World Bank and regional development banks like the Asian Development Bank and the African Development Bank focus on financing projects that contribute to long-term sustainable development. Microfinance institutions, on the other hand, provide small-scale financial services to entrepreneurs and low-income individuals who have limited access to traditional banking services. Agricultural banks specialize in offering credit and support to farmers and rural communities, while housing finance institutions facilitate affordable housing initiatives. Export-import banks assist in financing international trade and export activities, and industrial development corporations promote industrial growth through targeted funding and support (Xu, Ren, & Wu, 2019).

[1] College of Politics and Governance, Mahasarakham University, Thailand

[*] Corresponding Author; worrawoot.j@msu.ac.th

These specialized finance institutions are critical in addressing specific financing gaps and promoting economic growth in various sectors and regions. They help channel financial resources to underserved areas and sectors, enabling inclusive economic development. However, it is important to ensure that these institutions operate with transparency, good governance, and effective risk management to fulfill their intended objectives effectively.

In Thailand, the Bank for Agriculture and Agricultural Cooperatives (BAAC) is a significant financial institution entrusted with the responsibility of supporting farmers and alleviating poverty in the country. The bank's operations are subject to government control and regulation, aiming to address the challenges faced by Thai farmers, who constitute a substantial portion of the country's impoverished population (Bank for Agriculture and Agricultural Cooperatives, 2020). However, the bank encounters various issues and obstacles that hinder the achievement of its goals. This article will identify and discuss these challenges, with the ultimate aim of providing assistance and resolving the poverty issue in the country. The article will cover the following topics: 1) A Literature Review on Specialized Financial Institutions, 2) Understanding Thailand's Financial Institutions: A Brief Overview, 3) The BAAC and Its Policies Aimed at Assisting the Impoverished, 4) Key Challenges and Barriers to the Bank for Agriculture and Agricultural Cooperatives' Functioning, and 5) BAAC's Future: Strategic Plans and Operational Outlook.

**1. A Literature Review on Specialized Financial Institutions**

Numerous studies have focused on the involvement of governments in utilizing specialized financial institution mechanisms or tools to combat poverty among farmers and the underprivileged within a nation. In their 2020 study, Piskunov and Starikova (2020) investigated the challenges associated with government funding for infrastructure projects based on the public-private partnership model. The study identified major issues and proposed suggestions to enhance the effectiveness of support systems for public-private partnership projects in Russia. These suggestions included the establishment of dedicated specialized financial institutions, informed by international experiences and the current situation in Russia.

Siriwardane (2019) examined the impact of capital shocks on pricing in the CDS market. The study found that seller capital shocks, as assessed by CDS portfolio margin payments, accounted for a substantial 12 percent of the variation in weekly spread movements. This indicated that seller shocks provided unique information separate from institution-wide assessments, suggesting a segmented market where frictions among specialized financial institutions hindered capital entry in shorter time horizons. The investigation focused on regulating the terms of the creator, investor, treasurer, and director involved in investment funds.

Hasan (2019) explored investment funds as contemporary investment options that allow individuals lacking asset management skills to participate in local or international financial markets. Investment funds pool resources from multiple investors and are managed by specialized financial organizations to maximize returns. The study emphasized the regulation of agreements among the creator, investor, treasurer, and director in these funds.

Gomathi and Sugumar (2019) highlighted the importance of the agriculture sector as a major component of the economy and a foundation for overall national development. Developing nations like India face challenges in increasing agricultural productivity to meet the needs of growing populations. Specialized financial organizations such as NABARD and lead bank programs were implemented to enhance input utilization efficiency in agriculture. However, there still exists a disparity between the supply and demand for food grains produced in Tamil Nadu and India.

Shamsudin, Sarif, Mohammed, and Kamil (2018) discussed Development Financial Institutions (DFIs) as specialized financial institutions designated by the government to carry out socio-economic development tasks. The study focused on the involvement of Shari'ah committee members in determining the strategic direction of DFIs and assessing their alignment with Islamic economic goals. It recommended that DFIs aim their efforts towards achieving Islamic economic objectives and involve Shari'ah committee members in decision-making to ensure the honest achievement of these goals.

Lawal, Iyiola, and Adegbuyi (2018) emphasized the importance of entrepreneurship development in fostering economic prosperity, growth, and innovation. The study highlighted the significance of an efficient financial system in providing appropriate financial capital to entrepreneurs for growth and expansion. It examined the use of alternative finance sources and targeted government funding assistance among Nigerian entrepreneurs. The study found that alternative finance channels were increasingly being utilized alongside government assistance and specialized financial institutions to meet the financing needs of SMEs. It emphasized the necessity of well-intentioned government policies and frameworks for affordable financing to address entrepreneurial finance requirements.

Jouida, Bouzgarrou, and Hellara (2017) investigated the impact of activity and geographic diversification on the performance of financial institutions. The study analysed data from 412 French financial organizations over a ten-year period. It found a conflicting relationship between performance and diversification, but a positive link emerged when institutions employed a dual diversification strategy. The study proposed a taxonomy of French financial institutions and highlighted the importance of diversification for specialized financial institutions. The findings were consistent across performance and diversity metrics and considered potential endogeneity issues.

These research articles discussed various studies on specialized financial institutions and their roles in addressing poverty, infrastructure funding, agricultural development, Islamic economic goals, entrepreneurship, and financial sector diversification. The studies highlight challenges, propose improvements, and examine the impact of different factors on the performance of financial institutions. Key development ideas to study further include exploring the effectiveness of specialized financial institutions in poverty alleviation, improving government funding mechanisms for infrastructure projects, analysing the impact of capital shocks on financial markets, assessing the role of investment funds in facilitating market participation, examining strategies to enhance agricultural productivity, aligning development institutions with Islamic economic goals, studying the effectiveness of entrepreneurial support programs, and investigating the impact of diversification on financial institution performance. These areas offer promising

5opportunities for research and policy development to foster economic growth, financial inclusion, and sustainable development.

## 2. Understanding Thailand's Financial Institutions: A Brief Overview

Thailand boasts a diverse range of financial institutions, encompassing both public and private organizations. Financial institutions serve as entities engaged in lending activities or act as intermediaries between lenders and borrowers. Their revenue stems from the interest rate differential between what they receive from borrowers and what they pay to lenders. In Thailand, financial institutions can be classified into two main groups (Bank of Thailand, 2020b):

Firstly, Depositary Corporations, this group includes various entities such as commercial banks, Special Financial Institutions (SFIs), Savings Cooperatives, credit unions, and money market mutual funds. These institutions primarily engage in deposit-taking activities and provide financial services related to loans, savings, and investments. Secondly, Non-depository corporations: This category comprises entities like mutual funds, insurance companies, provident funds, asset management companies, and securities companies. Non-depository corporations primarily focus on offering investment and insurance products, managing assets, and facilitating securities transactions.

These classifications encompass the diverse landscape of financial institutions operating in Thailand, catering to different financial needs and services within the country. While Thailand boasts a diverse array of financial institutions, commercial banks stand out as the most prominent and widely recognized. Commercial banks hold this distinction due to their close proximity to the public and the extensive utilization of their services by individuals. Nonetheless, other financial institutions also fulfill specific functions within the financial landscape. Among them are special financial institutions, which serve as state agencies, operating as mechanisms to implement government policies in response to specific objectives.

Special financial institutions are established under specific laws with the primary goal of supporting government policies related to economic growth and investment promotion. The Bank for Agricultural and Agricultural Cooperatives (BAAC), Government Savings Bank (GSB), Government Housing Bank (GHB), and Export-Import Bank of Thailand (EXIM) are notable examples of such institutions. Each of these institutions has distinct objectives that are established in accordance with their respective mandates, as indicated in the table provided below (Bank of Thailand, 2020a).

**Table 1 presents key special financial institutions in Thailand**

| Financial institutions | Founded | Functions |
|---|---|---|
| Bank for Agriculture and Agricultural Cooperatives (BAAC) | 1966 | Provision of financial aid to support the livelihood of farmers and agricultural cooperatives who are in a weak financial condition. |
| Government Savings Bank (GSB) | 1913 | Encourages low-income consumption savings to make savings and then deposit savings to the bank. |
| Government Housing Bank (GH BANK) | 1953 | Providing financial aid to individuals to provide their own houses and helping the legal body to assign land and houses to individuals with low to medium incomes. |
| Export-Import Bank of Thailand (EXIM Thailand) | 1993 | Conduct undertakings that encourage and sustain Thai exports, imports and investments for national growth by offering credit facilities, guarantees, risk insurance, or other services that are conducive to achieving their objectives. |

Source: Bank of Thailand (2020a)

The primary focus of this article is to delve into the roles and functions of the BAAC. In order to comprehend its responsibility in alleviating poverty, particularly among farmers and the underprivileged, it is crucial to gain an understanding of the bank's structure, functions, policies, and current status.

### 3. The BAAC and Its Policies Aimed at Assisting the Impoverished

The Bank for Agriculture and Agricultural Cooperatives (BAAC) is a specialized financial institution owned by the state and operates under the Ministry of Finance. Its primary objective is to serve as a crucial government mechanism that supports rural development. The bank offers financial and developmental assistance to various target groups in rural areas, such as farmers, entrepreneurs, citizen groups, community organizations, and cooperatives, which form the foundation of Thailand's grassroots economy. Additionally, the government has assigned the bank with the important task of alleviating the challenges faced by farmers due to debt obligations and low prices of agricultural commodities (Bank for Agriculture and Agricultural Cooperatives, 2019; Fitchett, 1999).

BAAC has long played a pivotal role in Thailand's rural economy, particularly in supporting impoverished farmers and vulnerable communities. Founded in 1966, the BAAC's mission is to provide financial services tailored to the agricultural sector, focusing on poverty alleviation and improving the livelihoods of small-scale farmers. Its policies and programs are designed to address the economic challenges faced by low-income individuals while fostering sustainable agricultural development (BAAC, 2024a).

One of the core initiatives of BAAC is offering low-interest loans to farmers and small agricultural businesses. These loans are designed to provide much-needed capital for investments in seeds, equipment, livestock, and other essential resources. By easing access to credit, BAAC helps farmers increase productivity and income, allowing them to break the cycle of poverty. In addition to regular agricultural loans, BAAC offers specialized programs for those in the most vulnerable groups, including women, the elderly, and those in remote areas. These programs typically feature extended repayment periods and lower interest rates, tailored to the unique circumstances of these populations. This flexibility helps farmers focus on long-term sustainability rather than short-term financial burdens (BAAC, 2024b).

BAAC also promotes the formation of agricultural cooperatives, which allow small farmers to pool resources, share knowledge, and collectively bargain for better prices and market access. Through cooperatives, impoverished farmers can benefit from economies of scale, which in turn enhances their ability to compete in larger markets. BAAC provides financial support to these cooperatives through low-interest loans and grants, ensuring that they can continue to grow and offer crucial services to their members (BAAC, 2024b).

Recognizing the need for even smaller-scale financial services, BAAC has developed microfinance initiatives aimed at providing credit to rural individuals who may not qualify for traditional loans. These microfinance programs are particularly crucial for marginalized communities, offering them the opportunity to start small businesses or improve their agricultural productivity. The goal is to empower individuals to create their own sources of income, fostering self-reliance and reducing dependence on external aid (BAAC, 2024b).

In addition to financial assistance, BAAC's policies focus on promoting sustainable agricultural practices as part of its poverty alleviation strategy. Through its Bio Economy Credit, Circular Economy Credit, and Green Credit, BAAC encourages investment in technologies and practices that reduce waste, conserve natural resources, and promote organic farming. These initiatives help farmers adopt more efficient and environmentally friendly methods, ultimately improving yields and reducing long-term costs. By supporting sustainable practices, BAAC not only helps improve the economic stability of farmers but also ensures that future generations can continue to rely on agriculture as a viable means of livelihood. The focus on sustainable development aligns with the government's broader goals of transitioning Thailand toward an eco-friendlier economy (BAAC, 2024c).

BAAC's role in poverty alleviation extends beyond simply providing credit. The bank actively works to improve financial literacy among rural populations, teaching individuals how to manage their finances, plan for the future, and make informed decisions about borrowing and investment. Financial education is a critical component of BAAC's efforts to create lasting change, ensuring that farmers are better equipped to manage debt and navigate financial challenges (BAAC, 2024b).

The BAAC's policies aimed at assisting the impoverished are multifaceted, addressing not only immediate financial needs but also long-term sustainability and economic independence. Through targeted loan programs, support for cooperatives, microfinance initiatives, and the promotion of sustainable practices, BAAC plays a crucial role in reducing poverty in Thailand's agricultural sector. By empowering farmers and fostering a culture of self-reliance, BAAC continues to be a cornerstone of rural development in the country (BAAC, 2024b).

**4. Key Challenges and Barriers to the Bank for Agriculture and Agricultural Cooperatives' Functioning**

The Bank for Agriculture and Agricultural Cooperatives can be classified as a Microfinance Institution (MFI). Generally, the term "microfinance institutions" pertains to financial institutions that specifically focus on assisting underprivileged households and small businesses in accessing financial services (Greuning, Gallardo, & Randhawa, 1999; Hardy, Holden, & Prokopenko, 2002; Ledgerwood & White, 2006). Simultaneously, microfinance institutions (MFIs) have emerged as crucial lending institutions in the development process. Due to capital constraints, microfinance organizations have strived to attain varying degrees of sustainability (Bogan, 2012).

The Bank for Agriculture and Agricultural Cooperatives (BAAC) is recognized as a microfinance institution (MFI) in Thailand. Microfinance institutions are financial organizations designed to provide small-scale financial services, such as loans, savings accounts, and insurance, to underserved and underprivileged populations. The BAAC's primary focus is on rural households, particularly small farmers, agricultural workers, and small businesses that lack access to traditional banking services.

As an MFI, the BAAC provides financial services to low-income households and small-scale businesses that are often excluded from mainstream financial institutions. Traditional banks typically avoid lending to these groups due to their lack of collateral, credit history, or stable income. However, the BAAC's mission is to promote financial inclusion, ensuring that rural communities can access the financial support necessary to improve their livelihoods and contribute to economic development. BAAC's financial services are tailored to meet the unique needs of farmers and small businesses. The core of its offering is microcredit, which provides small loans to individuals and businesses that can be used for agricultural inputs, equipment, or small-scale entrepreneurial activities. The BAAC's loan schemes are designed to be flexible and affordable, featuring low-interest rates and extended repayment periods to alleviate the financial burden on borrowers (BACC, 2024).

In addition to microcredit, BAAC offers savings accounts that encourage rural communities to develop a habit of saving and building financial resilience. These accounts provide a secure way for individuals to manage their finances, save for emergencies, and invest in their future. BAAC also provides various insurance products, particularly crop insurance, to help farmers mitigate the financial risks associated with unpredictable weather and market conditions. The BAAC's primary target audience includes underprivileged households, small-scale farmers, and micro-enterprises that are the backbone of Thailand's rural economy. These groups often struggle to access financial services due to their geographic isolation, lack of formal education, or limited financial literacy. The BAAC steps in to fill this gap, offering financial products and services that are accessible, affordable, and designed to empower these communities. Small businesses in rural areas also benefit from BAAC's microfinance services. Many of these enterprises lack access to capital and struggle to grow due to limited funding. By providing microloans, the BAAC helps these businesses expand, purchase necessary equipment, hire additional workers, and contribute to the overall development of their local economies (BACC, 2024).

Beyond providing financial support, the BAAC is committed to promoting sustainable development. Many of its microfinance programs are aligned with broader environmental and social goals, such as encouraging organic farming, reducing waste, and adopting eco-friendly practices. Through initiatives like Bio Economy Credit and Green Credit, the BAAC supports investments in sustainable agriculture, ensuring that rural communities can improve their economic standing while conserving natural resources (BAAC, 2024c).

In its role as a microfinance institution, the BAAC serves as a crucial pillar for Thailand's rural development. By providing tailored financial services to underprivileged households and small businesses, it fosters financial inclusion, helps reduce poverty, and promotes long-term sustainability in the agricultural sector. Through its commitment to empowering rural communities, the BAAC continues to play a vital role in shaping Thailand's economic landscape.

**5. BAAC's Future: Strategic Plans and Operational Outlook**

The current and potential circumstances of the Bank for Agriculture and Agricultural Cooperatives (BAAC) as a state-owned enterprise are influenced by several factors. One major concern is its organizational structure. According to Section 14 of the BAAC Act (1966), the governance of the BAAC involves a Board of Directors composed of the Minister of Finance as Chairman, a Vice-Chairman, up to 12 additional Directors appointed by the Cabinet, and the Chairman of the BAAC serving as both Director and Secretary. This specific organizational arrangement presents challenges and could affect the effectiveness of the BAAC's operations (Bank for Agriculture and Agricultural Cooperatives, 2019). The primary issue causing policy failures at the Bank for Agriculture and Agricultural Cooperatives (BAAC) is the interference of politicians in its operations. To address this problem, it is essential to minimize political influence within the BAAC. This can be accomplished by improving the organization's knowledge and expertise and by implementing a comprehensive, long-term strategy. To effectively tackle the

11challenges faced by farmers and disadvantaged groups, the BAAC should integrate sustainable development principles into its policies and strategies. Doing so will enable the bank to better serve its target beneficiaries and foster enduring positive change.

Fortunately, the Bank for Agriculture and Agricultural Cooperatives (BAAC) has developed a potential roadmap for achieving sustainable development. According to their 2023 annual report, BAAC demonstrates a strong commitment to advancing the agricultural sector through a focus on exceptional customer service. Their strategy involves enhancing the capabilities of Thailand's agricultural sector by building on cooperative and entrepreneurial foundations. This approach includes using group networks to boost integrated manufacturing sectors, with a particular emphasis on empowering smart farmers. BAAC plans to introduce innovative technologies and practices that benefit both consumers and society at large. Additionally, the bank aims to enhance income opportunities and provide equitable access to investment funds for low-income individuals. By aligning its initiatives with the Sustainable Development Goals (SDGs), BAAC seeks to make a significant contribution to sustainable development and address key societal challenges (BAAC, 2024c).

The future operations of the Bank for Agriculture and Agricultural Cooperatives (BAAC) will depend on its ability to address and overcome current challenges and obstacles. To enhance its support for farmers and disadvantaged communities, the BAAC must implement strategies to guard against political manipulation and corruption. Improving transparency and accountability within the organization is essential. Furthermore, the BAAC should work on bolstering its expertise in agricultural and rural development to deliver more targeted and effective financial services. Engaging with relevant stakeholders, including government bodies and international organizations, will also play a key role in the BAAC's future success in combating poverty and fostering sustainable rural development. Ongoing evaluation and refinement of its policies and operations will be critical for the BAAC to achieve its mission effectively in the years ahead.

**Conclusion**

The Bank for Agriculture and Agricultural Cooperatives (BAAC) is a state-run financial institution specifically tasked with supporting farmers and disadvantaged groups within the country. Although the bank has achieved some success historically, it struggles with notable weaknesses and constraints arising from political interference and control. This lack of autonomy undermines its capacity to develop and execute effective strategies for public aid, leading to insufficient support for poverty alleviation efforts. To overcome these challenges, it is essential to employ professional methods that utilize technology and research tools, facilitating more efficient and targeted solutions for the issues faced by farmers and marginalized communities.